\documentstyle{article}
\newlength{\defaultparindent}
\newenvironment{Normal Times}{%
\Large %
\setlength{\parskip}{12pt}}{}
\begin{document}
%
\begin{center}
Forthcoming in Savitt, S. (ed), {\it Time's Arrows Today}, Cambridge
University Press, 1994.
\end{center}

\vskip 1.5cm
\begin{center}
{\Large {\bf {\it Cosmology, Time's Arrow, and That Old Double
Standard}}}
\vskip 1cm

{\Large Huw Price}

\vskip 1cm

School of Philosophy

University of Sydney

Australia 2006

[email: huw@extro.su.edu.au]

\end{center}

\vskip 2cm

\begin{center}{\Large Abstract}%
{\sc }

\end{center}
It has become widely accepted that temporal asymmetry is largely a
cosmological problem; the task of explaining temporal asymmetry reduces
in the main to that of explaining an aspect of the condition of the early
universe. However, cosmologists who discuss these issues often make
mistakes similar to those that plagued nineteenth century discussions of
the statistical foundations of thermodynamics. In particular, they are often
guilty of applying temporal "double standards" of various kinds---e.g., in
failing to recognise that certain statistical arguments apply with equal force
in either temporal direction.

This paper aims to clarify the issue as to what would count as adequate
explanation of cosmological time asymmetry. A particular concern is the
question whether it is possible to explain why entropy is low near the Big
Bang without showing that it must also be low near a Big Crunch, in the
event that the universe recollapses. I criticise some of the objections raised
to this possibility, showing that these too often depend on a temporal
double standard. I also discuss briefly some issues that arise if we take the
view seriously. (Could we observe a time-reversing future, for example?)

\newpage
A century or so ago, Ludwig Boltzmann and others attempted to
explain the temporal asymmetry of the second law of thermodynamics.
The hard-won lesson of that endeavour---a lesson still commonly
misunderstood---was that the real puzzle of thermodynamics lies
not in the question why entropy increases with time, but in that
as to why it was ever so low in the first place. To the extent
that Boltzmann himself appreciated that this was the real issue,
the best suggestion he had to offer was that the world as we
know it is simply a product of a chance fluctuation into a state
of very low entropy. (His statistical treatment of thermodynamics
implied that although such states are extremely improbable, they
are bound to occur occasionally, if the universe lasts a sufficiently
long time.) This is a rather desperate solution to the problem
of temporal asymmetry, however,%
{\normalsize %
\footnote{Not least of its problems is the fact that it implies
that all our historical evidence is almost certainly misleading:
for the `cheapest' or most probable fluctuation compatible with
our present experience will always be one which simply creates
the world as we find it, rather than having it evolve from an
earlier state of even lower entropy.}%
} and one of the great achievements of modern cosmology has been
to offer us an alternative. It now appears that temporal asymmetry
is cosmological in origin, a consequence of the fact that entropy
is much lower than its theoretical maximum in the region of the
Big Bang---i.e., in what we regard as the %
{\it early} stages of the universe.

\setlength{\parindent}{\defaultparindent}%
\setlength{\parskip}{0pt} The task of explaining temporal asymmetry
thus becomes the task of explaining this condition of the early
universe. In this paper I want to discuss some philosophical
constraints on the search for such an explanation. In particular,
I want to show that cosmologists who discuss these issues often
make mistakes which are strikingly reminiscent of those which
plagued the nineteenth century discussions of the statistical
foundations of thermodynamics. The most common mistake is to
fail to recognise that certain crucial arguments are blind to
temporal direction, so that any conclusion they yield with respect
to one temporal direction must apply with equal force with respect
to the other. Thus writers on thermodynamics often failed to
notice that the statistical arguments concerned are inherently
insensitive to temporal direction, and hence unable to account
for temporal asymmetry. And writers who did notice this mistake
commonly fell for another: recognising the need to justify the
double standard---the application of the arguments in question
`towards the future' but not `towards the past'---they appealed
to additional premisses, without noticing that in order to do
the job, these additions must effectively embody the very temporal
asymmetry which was problematic in the first place. To assume
the uncorrelated nature of initial particle motions (or incoming
`external influences'), for example, is simply to move the problem
from one place to another. (It may %
{\it look} less mysterious as a result, but this is no real
indication of progress. The fundamental lesson of these endeavours
is that much of what needs to be explained about temporal asymmetry
is so commonplace as to go almost unnoticed. In this area more
than most, folk intuition is a very poor guide to explanatory
priority.)

 One of the main tasks of this paper is to show that mistakes
of these kinds are widespread in modern cosmology, even in the
work of some of the contemporary physicists who have been most
concerned with the problem of the cosmological basis of temporal
asymmetry---in the course of the paper we shall encounter illicit
applications of a temporal double standard by Paul Davies, Stephen
Hawking and Roger Penrose, among others. Interdisciplinary point-
scoring
is not the primary aim, of course: by drawing attention to these
mistakes I hope to clarify the issue as to what would count as
adequate cosmological explanation of temporal asymmetry.

 I want to pay particular attention to the question as to whether
it is possible to explain why entropy is low near the Big Bang
without thereby demonstrating that it must be low near a Big
Crunch, in the event that the universe recollapses. The suggestion
that entropy might be low at both ends of the universe was made
by Thomas Gold in the early 1960s.%
{\normalsize %
\footnote{See Gold (1962) for example.}%
} With a few notable exceptions, cosmologists do not appear to
have taken Gold's hypothesis very seriously. Most appear to believe
that it leads to absurdities or inconsistencies of some kind.
However, I want to show that cosmologists interested in time
asymmetry continue to fail to appreciate how little scope there
is for an explanation of the low entropy Big Bang which does
not commit us to the Gold universe. I also want criticise some
of the objections that are raised to the Gold view, for these
too often depend on a temporal double standard. And I want to
discuss, briefly and rather speculatively, some issues that arise
if we take the view seriously. (Could we observe a time-reversing
future, for example?)

\setlength{\parindent}{\defaultparindent}%
\setlength{\parskip}{0pt} Let me begin with a very brief characterisation
of what it is about the early universe that needs to be explained.
There seems to be widespread agreement about this, so that if
what I say is not authoritative, neither is it particularly controversial.%
{\normalsize %
\footnote{There is an excellent account of this in Penrose (1989),
ch. 7.}%
} The question of the origin of the thermodynamic arrow appears
to come down to that as to why the early universe had just the
right degree of inhomogeneity to allow the formation of galaxies:
had it been more homogeneous, galaxies would never have formed;
had it been less homogeneous, most of the matter would have quickly
ended up in huge black holes. In either case we wouldn't have
galaxies as the powerhouses of most of the asymmetric phenomena
we are so familiar with. For present purposes the latter issue---that
as to why the universe is not less homogeneous---is the more
pressing. This is because, as we'll see, there are strong arguments
to the effect that if the universe recollapses, the other extremity
will be very inhomogeneous. So the homogeneity in the region
of the Big Bang would appear to represent a stark temporal asymmetry
in the universe as a whole.

\setlength{\parindent}{\defaultparindent}%
\setlength{\parskip}{0pt}%
{\it }

\begin{center}The natural mistake%
{\sc }

\end{center}The contemporary cosmological descendant of the
problem that Boltzmann was left with---the problem as to why
entropy was low in the past---is thus the question Why is the
universe so smooth near the Big Bang? Now in effect this question
is a call for an explanation of an observed feature of the physical
universe, and one common kind of response to a request for an
explanation of an observed state of affairs is to try to show
that `things had to be like that', or at least that it is in
some sense very probable that they should be like that. In other
words, it is to show that the state of affairs in question represents
the %
{\it natural} way for things to be. Accordingly, we find many
examples of cosmologists trying to show that the state of the
early universe is not really particularly special. For example,
the following remarks are from one of Paul Davies' popular accounts
of cosmology and time asymmetry: `It is clear that a time-asymmetric
universe does not demand any very special initial conditions.
It seems to imply a creation which is of a very general and random
character at the microscopic level. This initial randomness is
precisely what one would expect to emerge from a singularity
which is completely unpredictable.'%
{\normalsize %
\footnote{Davies (1977), 193-4.}%
}

\setlength{\parindent}{\defaultparindent}%
\setlength{\parskip}{0pt} The mistaken nature of this general
viewpoint has been ably pointed out by Roger Penrose, however.
Penrose asks what proportion of possible states of the early
universe---what percentage of the corresponding points in the
phase space of the universe---exhibit the degree of smoothness
apparent in the actual early universe. He gives a variety of
estimates, the most charitable of which (to the view that the
early universe is `natural') allows that as many as 1 in 10%
{\normalsize $^{10^{30}}$}%
 possible early universes are like this!%
{\normalsize %
\footnote{His most recent estimate is 1 in 10$^{10^{120}}$; see Penrose
(1989), ch. 7.}%
}

\setlength{\parindent}{\defaultparindent}%
\setlength{\parskip}{0pt} There is another way to counter the
suggestion that the smooth early universe is statistically natural,
an argument more closely related to our central concerns in this
paper. It is to note that we would not regard a collapse to a
smooth %
{\it late} universe (just before a Big Crunch) as statistically
natural---quite the contrary, as we noted above---but that in
the absence of any prior reason for thinking otherwise, this
consideration applies just as much to one end of the universe
as to the other. In these statistical terms, then, a smooth Big
Bang should seem just as unlikely as a smooth Big Crunch.%
{\normalsize %
\footnote{Penrose himself makes a point of this kind; see Penrose
(1989), 339. However, we shall see that Penrose fails to appreciate
the contrapositive point, which is that if we take statistical
reasoning to be inappropriate towards the past, we should not
take it to be appropriate with respect to the future. }%
}

\setlength{\parindent}{\defaultparindent}%
\setlength{\parskip}{0pt} The fact that this simple argument
goes unnoticed reflects the difficulty that we have in avoiding
the double standard fallacies in these cases. It deserves more
attention, however. Indeed, its importance goes beyond its ability
to counter the tendency to regard a smooth early universe as
`natural', and hence not in need of explanation. As we shall
see, it also defuses the most influential arguments against Gold
models of the universe, in which entropy eventually decreases.
It is common to find considerations concerning gravitational
collapse offered as decisive objections the Gold view. However,
what the objectors fail to see is that if the argument were decisive
in one temporal direction it would also be decisive in the other,
in which case its conclusion would conflict with what we know
to be the case, namely that entropy decreases towards the Big
Bang.

 In view of the importance of the argument it is worth spelling
it out in more detail, and giving it a name. I'll call it the
{\it Gravitational Argument from Symmetry} (GAS, for short).
The argument has three main steps:

\setlength{\parindent}{-14pt}%
\begin{list}{ }{%
\setlength{\leftmargin}{15pt}\setlength{\rightmargin}{0pt}%
\setlength{\topsep}{0pt}\setlength{\partopsep}{0pt}}
\item
\item 1.
We consider the natural condition of a universe at the end of
a process of gravitational collapse---in other words, we ask
what the universe might be expected to be like in its late stages,
when it collapses under its own weight. As noted above, the answer
is that it is overwhelmingly likely to be in a very inhomogeneous
state---clumpy, rather than smooth.

2. We reflect on the fact that if we view the history of our
universe in reverse, what we see is a universe collapsing under
its own gravity, accelerating towards a Big Crunch. As argued
in step 1, the natural destiny for such a universe is not the
smooth one we know our own universe to have.

3. We note that there is no objective sense in which this reverse
way of viewing the universe is any less valid than the usual
way of viewing it. Nothing in physics tells us that there is
a wrong or a right way to choose the orientation of the temporal
coordinates. %
{\it Nothing in physics tells us that one end of the universe
is objectively the start and the other end objectively the finish}.
In other words, the perspective adopted at step 2 is just as
valid as a basis for determining the natural condition of what
we normally call the early universe as the standard perspective
is for determining the likely condition of what normally call
the late universe.

\item
\end{list}%
\setlength{\parindent}{\defaultparindent}

 The lesson of GAS is that there is much less scope for differentiating
the early and late stages of a universe than tends to be assumed.
If we want to treat the late stages in terms of a theory of gravitational
collapse, we should be prepared to treat the early stages in
the same way. Or in other words if we treat the early stages
in some other way---in terms of some additional boundary constraint,
for example---then we should be prepared to consider the possibility
that the late stages may be subject to the same constraint. Failure
to appreciate this point has tended to obscure what may justly
be called the %
{\it basic dilemma} of cosmology and temporal asymmetry. In
virtue of the symmetry considerations, it seems that our choices
are either to follow Gold, admitting reversal of the thermodynamic
arrow in the case of gravitational collapse; or to acknowledge
that temporal asymmetry is simply inexplicable---i.e., that the
low initial entropy of the universe is not a predictable consequence
of our best physical theories of the universe as a whole.

 Later on I shall discuss a range of possible responses to this
basic dilemma. First of all, however, in order to illustrate
how thoroughly contemporary cosmologists have failed to appreciate
the dilemma and the symmetry considerations on which it rests,
I want to discuss two recent suggestions as to the origins of
cosmological time asymmetry.

{\it }

\begin{center}The appeal to inflation

\end{center}The first case stems from the inflationary model.
The basic idea here is that in its extremely early stages the
universe undergoes a period of exponential expansion, driven
by a gravitational force which at that time is repulsive rather
than attractive. At the end of that period, as gravity becomes
attractive, the universe settles into the more sedate expansion
of the classical Big Bang.%
{\normalsize %
\footnote{For a general introduction to the inflationary model
see Lind\'{e} (1987).}%
} Among the attractions of this model has been thought to be
its ability to account for the smoothness of the universe after
the Big Bang. However, the argument for this conclusion is essentially
a statistical one: the crucial claim is that the repulsive gravity
in the inflationary phase will tend to `iron out' inhomogeneities,
leaving a smooth universe at the time of the transition to the
classical Big Bang. The argument is presented by Paul Davies,
who concludes that `the Universe ... began in an arbitrary, rather
than remarkably specific, state. This is precisely what one would
expect if the Universe is to be explained as a spontaneous random
quantum fluctuation from nothing.'%
{\normalsize %
\footnote{Davies (1983), 398.}%
}

\setlength{\parindent}{\defaultparindent}%
\setlength{\parskip}{0pt} This argument graphically illustrates
the temporal double standard that commonly applies in discussions
of these problems. The point is that as in step 2 of GAS we might
equally well argue, viewing the expansion from the Big Bang in
reverse, that (what will then appear as) the gravitational %
{\it collapse} to the Big Bang must produce %
{\it inhomogeneities} at the time of the transition to the inflationary
phase (which will now appear as a deflationary phase, of course).
{\it Unless one temporal direction is already privileged, the
argument is as good in one direction as the other}. So in the
absence of a justification for the double standard---a reason
to apply the statistical argument in one direction rather than
the other---Davies' argument cannot possibly do the work required
of it.

 This is close to the point made in reply to Davies by Don Page.%
{\normalsize %
\footnote{Page (1983).}%
} Page objects that in arguing statistically with respect to
behaviour during the inflationary phase, Davies is in effect
assuming the very time asymmetry which needs to be explained
(i.e., that entropy increases). However, Davies might reply that
statistical reasoning is acceptable in the absence of constraining
boundary conditions. It seems to me there are two possible replies
at this point. One might argue (as Page does) that initial conditions
have to be special to give rise to inflation in the first place,
and hence that Davies' imagined initial conditions are in fact
far from arbitrary. Or more directly one might argue as I have,
viz. that if there is no boundary constraint at the time of transition
from inflationary phase to classical Big Bang, then we are equally
entitled to argue from the other direction, with the conclusion
that the universe is inhomogeneous at this stage.

\setlength{\parindent}{\defaultparindent}%
\setlength{\parskip}{0pt} Davies himself argues that `a recontracting
Universe arriving at the big crunch would not undergo ``deflation'',
for this would require an exceedingly improbable conspiracy of
quantum coherence to reverse-tunnel through the phase transition.
There is thus a distinct and fundamental asymmetry between the
beginning and the end of a recontracting Universe.'%
{\normalsize %
\footnote{Davies (1983), 399.}%
} However, he fails to notice that this is in conflict with the
argument he has given us concerning the other end of the universe,
a conflict which can only be resolved either (i) by acknowledging
that inflation is abnormal (and hence requires explanation, if
we are to explain temporal asymmetry); or (ii) by arguing that
although the coherence required for deflation looks improbable,
it is in fact guaranteed by the reverse of the argument Davies
himself uses with respect to the other end of the universe. But
to accept (ii) would be to accept that entropy decreases as the
universe recollapses, a view that as we shall see, Davies feels
can be dismissed on other grounds. As it is, therefore, Davies
is vulnerable to the charge that his own admission that collapse
doesn't require deflation automatically entails that expansion
doesn't require inflation. Again, this follows immediately from
the realisation that there is nothing objective about the temporal
orientation. A universe that collapses without deflation just
{\it is} a universe that expands without inflation. It is exactly
the same universe, under a different but equally valid description.%
{\normalsize %
\footnote{All the same, it might seem that there is an unresolved
puzzle here: as we approach the transition between an inflationary
phase and the classical phase from one side, most paths through
phase space seem to imply a smooth state at the transition. As
we approach it from the other side most paths through phase space
appear to imply a very non-smooth state. How can these facts
can be compatible with one another? I take it that the answer
is that the existence of the inflationary phase is in fact a
very strong boundary constraint, invalidating the usual statistical
reasoning from the `future' side of the transition. }%
}

\setlength{\parindent}{\defaultparindent}%
\setlength{\parskip}{0pt}%
{\it }

\begin{center}Hawking and the  Big Crunch%
{\normalsize %
\footnote{This section expands on some concerns expressed in
Price (1989).}%
}

\end{center}%
\setlength{\parindent}{\defaultparindent}%
\setlength{\parskip}{0pt}Our second example is better known,
having been described in Stephen Hawking's best seller, %
{\it A Brief History of Time}.%
{\normalsize %
\footnote{Hawking (1988).}%
} It is Hawking's proposal to account for temporal asymmetry
in terms of what he calls the %
{\it No Boundary Condition }(NBC)---a proposal concerning the
quantum wave function of the universe. To see what is puzzling
about Hawking's claim, let us keep in mind the basic dilemma.
It seemed that provided we avoid double standard fallacies, any
argument for the smoothness of the universe would apply at both
ends or at neither. So our choices seemed to be to accept the
globally symmetric Gold universe, or to resign ourselves to the
fact that temporal asymmetry is not explicable (without additional
assumptions or boundary conditions) by a time-symmetric physics.
The dilemma is particularly acute for Hawking, because he has
a more reason than most to avoid resorting to additional boundary
conditions. They conflict with the spirit of his NBC, namely
that one restrict possible histories for the universe to those
that `are finite in extent but have no boundaries, edges, or
singularities.'%
{\normalsize %
\footnote{Hawking (1988), 148.}%
}

\setlength{\parindent}{\defaultparindent}%
\setlength{\parskip}{0pt} Hawking tells us how initially he
thought that this proposal favoured the former horn of the above
dilemma: `I thought at first that the no boundary condition did
indeed imply that disorder would decrease in the contracting
phase.'%
{\normalsize %
\footnote{Hawking (1988), 150.}%
} He changed his mind, however, in response to objections from
two colleagues: `I realized that I had made a mistake: the no
boundary condition implied that disorder would in fact continue
to increase during the contraction. The thermodynamic and psychological
arrows of time would not reverse when the universe begins to
contract or inside black holes.'%
{\normalsize %
\footnote{Hawking (1988), 150.}%
}

\setlength{\parindent}{\defaultparindent}%
\setlength{\parskip}{0pt} This change of mind enables Hawking
to avoid the apparent difficulties associated with reversing
the thermodynamic arrow of time. What is not clear is how he
avoids the alternative difficulties associated with %
{\it not} reversing the thermodynamic arrow of time. That is,
Hawking does not explain how his proposal can imply that entropy
is low near the Big Bang, without equally implying that it is
low near the Big Crunch. The problem is to get a temporally asymmetric
consequence from a symmetric physical theory. Hawking suggests
that he has done it, but doesn't explain how. Readers are entitled
to feel a little dissatisfied. As it stands, Hawking's account
reads a bit like a suicide verdict on a man who has been stabbed
in the back: not an impossible feat, perhaps, but we'd like to
know how it was done!

 It seems to me that there are three possible resolutions of
this mystery. The first, obviously, is that Hawking has found
a way round the difficulty. The easiest way to get an idea of
what he would have to have established is to think of three classes
of possible universes: those which are smooth and ordered at
both temporal extremities, those which are ordered at one extremity
but disordered at the other, and those which are disordered at
both extremities. If Hawking is right, then he has found a way
to exclude the last class, without thereby excluding the second
class. In other words, he has found a way to exclude disorder
at one temporal extremity of the universe, without excluding
disorder at both extremities. Why is this combination the important
one? Because if we can't exclude universes with disorder at both
extremities, then we haven't explained why our universe doesn't
have disorder at both extremities---we know that it has order
at least one temporal extremity, namely the extremity we think
of as at the beginning of time. And if we do exclude disorder
at both extremities, we are back to the answer that Hawking gave
up, namely that order will increase when the universe contracts.

 Has Hawking shown that the second class of universal histories,
the order--disorder universes, are overwhelmingly probable? It
is important to appreciate that this would not be incompatible
with the underlying temporal symmetry of the physical theories
concerned. A symmetric physical theory might be such that all
or most of its possible realisations were asymmetric. Thus Hawking
might have succeeded in showing that the NBC implies that any
(or almost any) possible history for the universe is of this
globally asymmetric kind. If so, however, then he hasn't yet
explained to his lay readers how he managed it. In a moment I'll
describe my attempts to find a solution in Hawking's technical
papers. What seems clear is that it can't be done by reflecting
on the consequences of the NBC for the state of one temporal
extremity of the universe, considered in isolation. For if that
worked for the `initial' state it would also work for the `final'
state; unless of course the argument had illicitly assumed an
objective distinction between initial state and final state,
and hence applied some constraint to the former that it didn't
apply to the latter. What Hawking needs is a more general argument,
to the effect that disorder--disorder universes are impossible
(or at least overwhelmingly improbable). It needs to be shown
that almost all possible universes have at at least one ordered
temporal extremity---or equivalently, at most one disordered
extremity. (As Hawking points out, it will then be quite legitimate
to invoke a weak anthropic argument to explain why we regard
the ordered extremity thus guaranteed as an %
{\it initial} extremity. In virtue of its consequences for temporal
asymmetry elsewhere in the universe, conscious observers are
bound to regard this state of order as lying in their past.)

 That's the first possibility: Hawking has such an argument,
but hasn't told us what it is (probably because he doesn't see
why it is so important).%
{\normalsize %
\footnote{This loophole may be smaller than it looks. Hawking's
NBC would not provide an interesting explanation of temporal
asymmetry if it simply operated like the assumption that all
allowable models of the universe display the required asymmetry.
This would amount to putting the asymmetry in `by hand' (as physicists
say), to stipulating what we wanted to explain. If the NBC is
to exploit this loophole, in other words, it must imply this
asymmetry, while being sufficiently removed from it so as not
to seem ad hoc. }%
} As I see it, the other possibilities are that Hawking has made
one of two mistakes (neither of them the mistake he claims to
have made). Either his NBC does exclude disorder at both temporal
extremities of the universe, in which case his mistake was to
change his mind about contraction leading to decreasing entropy;
or the proposal doesn't exclude disorder at either temporal extremity
of the universe, in which case his mistake is to think that the
NBC accounts for the low entropy Big Bang.

\setlength{\parindent}{\defaultparindent}%
\setlength{\parskip}{0pt} I have done my best to examine Hawking's
published papers in order to discover which of these three possibilities
best fits the case. A helpful recent paper is Hawking's contribution
to a meeting on the arrow of time held in Spain in 1991.%
{\normalsize %
\footnote{Hawking (1993), Page references here are to a preprint
version.}%
} In this paper Hawking describes the process by which he and
various colleagues applied the NBC to the question of temporal
asymmetry. He recounts how he and Halliwell `calculated the spectrum
of perturbations predicted by the no boundary condition'.%
{\normalsize %
\footnote{Hawking (1993), 4.}%
} The conclusion was that `one gets an arrow of time. The universe
is nearly homogeneous and isotropic when it is small. But it
is more irregular, when it is large. In other words, disorder
increases, as the universe expands.'%
{\normalsize %
\footnote{Hawking (1993), 4.}%
} I want to note in particular that at this stage Hawking doesn't
refer to the stage of the universe in temporal terms---%
{\it start} and %
{\it finish}, for example---but only in terms of its size. Indeed
he correctly points out that the temporal perspective comes from
us, and depends in practice on the thermodynamic arrow.

\setlength{\parindent}{\defaultparindent}%
\setlength{\parskip}{0pt} Hawking then tells us how he

\setlength{\parindent}{-35.45pt}%
\begin{list}{ }{%
\setlength{\leftmargin}{35.45pt}\setlength{\rightmargin}{0pt}%
\setlength{\topsep}{0pt}\setlength{\partopsep}{0pt}}
\item

 made what I now realize was a great mistake. I thought that
the No Boundary Condition would imply that the perturbations
would be small whenever the radius of the universe was small.
That is, the perturbations would be small not only in the early
stages of the expansion, but also in the late stages of a universe
that collapsed again. ... This would mean that disorder would
increase during the expansion, but decrease again during the
contraction.

\end{list}%
\setlength{\parindent}{\defaultparindent}He goes on to say how
he was persuaded that this was a mistake, as a result of objections
raised by Page and Laflamme. He came to accept that

\setlength{\parindent}{-35.45pt}%
\begin{list}{ }{%
\setlength{\leftmargin}{35.45pt}\setlength{\rightmargin}{0pt}%
\setlength{\topsep}{0pt}\setlength{\partopsep}{0pt}}
\item

 When the radius of the universe is small, there are two kinds
of solution. One would be an almost euclidean complex solution,
that started like the north pole of a sphere, and expanded monotonically
up to a given radius. This would correspond to the start of the
expansion. But the end of the contraction would correspond to
a solution that started in a similar way, but then had a long
almost lorentzian period of expansion, followed by a contraction
to the given radius. ... This would mean that the perturbations
would be small at one end of time, but could be large and non
linear at the other end. So disorder and irregularity would increase
during the expansion, and would continue to increase during the
contraction.%
{\normalsize %
\footnote{Hawking (1993), 6.}%
}%
{\normalsize  }

\end{list}%
\setlength{\parindent}{\defaultparindent}%
\setlength{\parskip}{0pt}Hawking then describes how he and Glenn
Lyons have `studied how the arrow of time manifests in the various
perturbation modes'. He says that there are two relevant kinds
of perturbation mode, those that oscillate and those that don't.
The former `will be essentially time symmetric, about the time
of maximum expansion. In other words the amplitude of perturbation
will be the same at a given radius during the expansion as at
the same radius during the contraction phase.'%
{\normalsize %
\footnote{Hawking (1993), 6.}%
} The latter, by contrast, `will grow in amplitude in general.
They will be small when they come within the horizon during expansion.
But they will grow during the expansion, and continue to grow
during the contraction. Eventually they will become non linear.
At this stage, the trajectories will spread out over a large
region of phase space.'%
{\normalsize %
\footnote{Hawking (1993), 6-7.}%
} It is the latter perturbation modes which, in virtue of the
fact they are so much more common, lead to the conclusion that
disorder increases as the universe recontracts.

\setlength{\parindent}{\defaultparindent}%
\setlength{\parskip}{0pt} Let us focus on the last quotation.
If it is not an objective matter which end of the universe represents
expansion and which contraction, and there isn't a constraint
which operates simply in virtue of the radius of the universe,
why should the perturbations ever be small? Why can't they be
large at both ends, compatibly with the NBC?

 I have been unable to find an answer to this crucial question
in Hawking's papers. However, in an important earlier paper by
Hawking and Halliwell,%
{\normalsize %
\footnote{Hawking and Halliwell (1985).}%
} the authors consistently talk of showing that the relevant
modes %
{\it start off} in a particular condition. Let me give you some
examples (with my italics throughout). First from the abstract:
`We ... show ... that the inhomogeneous or anisotropic modes
{\it start off} in their ground state.'%
{\normalsize %
\footnote{Hawking and Halliwell (1985), 1977.}%
} Next from the introduction:

\setlength{\parindent}{-35.45pt}%
\setlength{\parskip}{0pt}%
\begin{list}{ }{%
\setlength{\leftmargin}{35.45pt}\setlength{\rightmargin}{0pt}%
\setlength{\topsep}{0pt}\setlength{\partopsep}{0pt}}
\item

 We show that the gravitational-wave and density-perturbation
modes obey decoupled time-dependent Schr\"{o}dinger equations
with respect to the time parameter of the classical solution.
The boundary conditions imply that these modes %
{\it start off} in the ground state. ...

 We use the path-integral expression for the wave function in
sec. VII to show that the perturbation wave functions %
{\it start out} in their ground states.%
{\normalsize %
\footnote{Hawking and Halliwell (1985), 1978.}%
}

\end{list}%
\setlength{\parindent}{\defaultparindent}%
\setlength{\parskip}{0pt} How are we to interpret these references
to how the universe %
{\it starts off}, or %
{\it starts out}? Do they embody an assumption that one temporal
extremity of the universe is objectively its start? Presumably
Hawking and Halliwell would want to deny that they do so, for
otherwise they have simply helped themselves to a temporal asymmetry
at this crucial stage of the argument. (As I have noted, Hawking
is in other places quite clear that our usual tendency to regard
one end of the universe as the start is anthropocentric in origin,
though related to the thermodynamic arrow---in virtue of their
dependence on the entropy gradient, sentient creatures are bound
to regard the low entropy direction as the past.) But without
this assumption what is the objective content of Hawking and
Halliwell's conclusions? Surely it can only be that the specified
results obtain when the universe is small; in which case the
argument works either at both ends or at neither.

 We have seen that in reconsidering his earlier views on the
fate of a collapsing universe, Hawking appears to be moved by
what is essentially a statistical consideration: the fact that
(as Page convinced him) most possible histories lead to a disordered
collapse. However, the lesson of GAS was that in the absence
of any prior justification for a temporal double standard, statistical
arguments defer to boundary conditions. (The fact that the Big
Bang is smooth trumps any general appeal to the clumpiness of
the end states of gravitational collapse.) Accordingly, it would
apparently have been open to Hawking to argue that Page's statistical
considerations were simply overridden by the NBC, treated as
a symmetric constraint on the temporal extremities of the universe.
Given that he does not argue this way, however, he needs to explain
why analogous statistical arguments do not apply towards (what
we call) the Big Bang.

 Thus it is my impression that Hawking did indeed make a mistake
about temporal asymmetry, but not the mistake he thought he made.
Either the mistake occurred earlier, if he and Halliwell have
illicitly relied on the assumption that the universe has an objective
start; or it occurred when Hawking failed to see that in virtue
of the fact that the Hawking-Halliwell argument depends only
on %
{\it size}, it does yield boundary conditions for both temporal
extremities of the universe, and hence excludes the possible
histories which Page and Laflamme took to be problematic for
his original endorsement of the time-reversing collapse. And
if he didn't make either of these mistakes, then at the very
least he has failed to see the need to avoid them, for otherwise
he would surely have explained how he managed to do so.

{\it }

\begin{center}The basic dilemma, and some ways to avoid it

\end{center}The above examples suggest that even some of the
most capable of modern cosmologists appear to have difficulty
in grasping what I called the basic dilemma of cosmology and
time asymmetry. To restate this dilemma, it is that apparently
have to accept either (option 1) that entropy decreases towards
all singularities, including future ones; or (option 2) that
temporal asymmetry, and particularly the low entropy condition
of the Big Bang, is not explicable by a time-symmetric physics.%
{\normalsize %
\footnote{It might be thought that the dilemma only arises if
the universe recollapses. Perhaps the universe is bound to be
such that it doesn't recollapse. (This is predicted by some inflationary
models.) But even if the universe as a whole never recollapses
to a singularity, it appears that parts of it do, as certain
massive objects collapse to black holes. The dilemma arises again
with respect to these regions. This point is made for example
by  Penrose (1979), 597-8.}%
}

\setlength{\parindent}{\defaultparindent}%
\setlength{\parskip}{0pt} The writer who has done most to bring
to our attention the force of this dilemma is Roger Penrose,
who chooses option 2.%
{\normalsize %
\footnote{See Penrose (1979), and particularly Penrose (1989),
ch. 7.}%
} (We shall look later at his reasons for rejecting option 1.)
Accordingly, Penrose suggests that there is an additional asymmetric
law of nature to the effect that initial singularities obey what
amounts to a smoothness constraint. In effect, he is arguing
that it is reasonable to believe that such a constraint exists,
because otherwise the universe as we find it would be wildly
unlikely. Penrose's use of the term %
{\it initial singularity} might itself seem to violate our requirement
that the terms %
{\it initial} and %
{\it final} not be regarded as of any objective significance.
However in this case the difficulty is clearly superficial: Penrose's
claim need only be that it is a physical law that there is one
temporal direction in which the Weyl curvature always approaches
zero towards singularities. The fact that conscious observers
inevitably regard that direction as the past will then follow
from the sort of weak anthropic argument we have already mentioned.%
{\normalsize %
\footnote{It seems to me that Penrose himself misses this point,
however. For example in discussing the Weyl curvature hypothesis
(Penrose (1989), 352-3), he gives us absolutely no indication
that he regards the fact that some singularities are initial
and others final as of anything other than objective significance.}%
}

\setlength{\parindent}{\defaultparindent}%
\setlength{\parskip}{0pt} Notice however that even an advocate
of the time-symmetric option 1 might be convinced by Penrose's
argument that the observed condition of the universe can only
be accounted for by an independent physical law---the difference
would be that such a person would take the law to be that the
Weyl curvature approaches zero towards %
{\it all} singularities. Before we turn to the various reasons
that have been offered for rejecting option 1, let me mention
some possible strategies towards the third option, which is somehow
to evade the dilemma altogether.

 The first possibility is the one mentioned above in our discussion
of Hawking's proposal. We noted that it is possible that a symmetric
theory might have only (or mostly) asymmetric models. This may
seem an attractive solution, at least in comparison to the alternatives,
but two notes of caution seem in order.

 First, as was pointed out earlier,%
{\normalsize %
\footnote{See footnote 17.}%
} a proposal of this kind needs to distance itself from the accusation
that it simply puts in the required asymmetry `by hand'. It is
difficult to lay down precise guidelines for avoiding this mistake---after
all, if the required asymmetry weren't already implicit in the
theoretical premisses in some sense, it couldn't be derived from
them---but presumably the intention should be that asymmetry
should flow from principles which are not %
{\it manifestly} asymmetric, and which have independent theoretical
justification.

\setlength{\parindent}{\defaultparindent}%
\setlength{\parskip}{0pt} Second, we should not be misled into
{\it expecting} a solution of this kind---predominantly asymmetric
models of a symmetric theory---by the sort of statistical reasoning
employed in step 1 of GAS. In particular, we should not think
that the intuition that the most likely fate for the universe
is a clumpy gravitational collapse makes a solution of this kind
prima facie more plausible than a globally symmetric model of
Gold's kind. The point of GAS was that these statistical grounds
are temporally symmetric: if they excluded Gold's suggestion
then they would also exclude models with a low entropy Big Bang.
In effect, the hypothesis that the Big Bang is explicable is
the hypothesis that something---perhaps a boundary condition
of some kind, perhaps an alternative statistical argument, conducted
in terms of possible models of a theory---defeats these statistical
considerations in cosmology, with the result that we are left
with no reason to %
{\it expect} an asymmetric solution in preference to Gold's
symmetric proposal. On the contrary, the right way to reason
seems to be something like this: the smoothness of the Big Bang
shows that statistical arguments based on the character of gravitational
collapse are not always reliable---on the contrary, they are
unreliable in the one case (out of a possible two!) in which
we can actually subject them to observational test. Having discovered
this, should we continue to regard them as reliable in the remaining
case (i.e., when oriented towards the Big Crunch)? Obviously
not, at least in the absence of any independent reason for applying
such a double standard.

 It is true that things would be different if we were prepared
to allow that the low entropy Big Bang is %
{\it not} explicable---that it is just a statistical `fluke'.
In this case we might well argue that we have very good grounds
to expect the universe to be `flukey' only at one end. However,
at this point we would have abandoned the strategy of trying
to show that almost all possible universes are asymmetric, the
goal of which was precisely to %
{\it explain} the low entropy Big Bang. Instead we might be
pursuing a different strategy altogether. Perhaps the reason
that the universe looks so unusual to us is simply that we can
only exist in very unusual bits of it. Given that we depend on
the entropy gradient, in other words, perhaps this explains why
we find ourselves in a region of the universe exhibiting such
a gradient.

 I am not going to explore this anthropic idea in any detail
here. Let me simply mention two large difficulties that it faces.
The first is that it depends on there being a genuine multiplicity
of actual `bits' of a much larger universe, of which our bit
is simply some small corner. It is no use relying on other merely
possible worlds.%
{\normalsize %
\footnote{Unless in David Lewis's sense, so that `actual' is
simply indexical, and denotes no special objective status. See
Lewis (1986).}%
} So the anthropic solution is exceedingly costly in ontological
terms. (This would not matter if the cost was one we were committed
to bearing anyway, of course, as perhaps in some inflationary
pictures, where universes in our sense are merely bubbles in
some grand foam of universes.)

\setlength{\parindent}{\defaultparindent}%
\setlength{\parskip}{0pt} The second difficulty is that as Penrose
emphasises,%
{\normalsize %
\footnote{Penrose (1979), 634.}%
} there may well be much less costly ways to generate a sufficient
entropy gradient to support life. Penrose argues that the observed
universe is still vastly more unlikely than life requires. However,
it is not clear that inflation does not leave a loophole here,
too. If the inflationary model could show that a universe of
the size of ours is an `all or nothing' matter, then the anthropic
argument would be back on track. The quantum preconditions for
inflation might be extremely rare, but this doesn't matter, so
long as (a) there is enough time in some background grand universe
for them to occur eventually and (b) when (and only when) they
do occur a universe of our sort arises, complete with its smooth
boundary. Hence it seems to me that this anthropic strategy is
still viable---albeit repugnant to well brought up Occamists!

\begin{center}%
\setlength{\parindent}{\defaultparindent}%
\setlength{\parskip}{0pt}The case against the Gold universe

\end{center}To choose the first horn of the basic dilemma is
to allow that our universe might have low entropy at both ends,
or more generally in the region of any singularity. Of course,
the issue then arises as to why this should be the case. One
option, perhaps in the end the only one, is to accept the low
entropy of singularities as an additional law of nature. As we
saw, this would be in the spirit of Penrose's proposal, but it
would still be a time-symmetric law. True, we might find such
a law somewhat %
{\it ad hoc}. But this might seem a price worth paying, if the
alternative is that we have no explanation for such a striking
physical anomaly.%
{\normalsize %
\footnote{These considerations apply as much to Penrose's asymmetric
proposal as to its symmetric variant. Note also that the proposal
might seem less ad hoc if attractively grounded on formal theoretical
considerations of some kind, as might arguably be true of Hawking's
NBC.}%
}

\setlength{\parindent}{\defaultparindent}%
\setlength{\parskip}{0pt} In suggesting the symmetric time-reversing
universe in the 1960s, Gold was attracted to the idea that the
expansion of the universe might account for the second law of
thermodynamics. He saw that this would entail that the law would
change direction if the universe recontracts. Thus the universe
would enter an age of apparent miracles. Radiation would converge
on stars, apples would compose themselves in decompost heaps
and leap into trees, and humanoids would arise from their own
ashes, grow younger, and become unborn. These humanoids wouldn't
see things this way, of course. Their psychological time sense
would also be reversed, so that from their point of view their
world would look much as ours does to us.

 However, by now it should be obvious that such apparently miraculous
behaviour cannot in itself constitute an objection to this symmetric
model of the universe, for reasons exactly analogous to those
invoked in GAS. Perhaps surprisingly (in view of his tendency
to appeal to a double standard elsewhere) this point is well
made by Davies. After describing some `miraculous' behaviour
of this kind, Davies continues: `It is curious that this seems
so laughable, because it is simply a description of our present
world given in reversed-time language. Its occurrence is %
{\it no more remarkable} than what we at present experience---indeed
it %
{\it is} what we actually experience---the difference in description
being purely semantic and not physical.'%
{\normalsize %
\footnote{Davies (1977), 196.}%
} Davies goes on to point out that the difficulty really lies
in managing the transition: `What %
{\it is} remarkable, however, is the fact that our `forward'
time world %
{\it changes into} [a] backward time world (or vice versa, as
the situation is perfectly symmetric).'

\setlength{\parindent}{\defaultparindent}%
\setlength{\parskip}{0pt} What exactly are the problems about
this transition? In the informal work from which I have just
quoted, Davies suggests that the main problem is that it requires
that the universe have very special initial conditions.

\begin{list}{ }{%
\setlength{\leftmargin}{35pt}\setlength{\rightmargin}{17.05pt}%
\setlength{\topsep}{0pt}\setlength{\partopsep}{0pt}}
\item Although
the vast majority of microscopic motions in the big bang give
rise to purely entropy-increasing worlds, a very, very special
set of motions could indeed result in an initial entropy increase,
followed by a subsequent decrease. For this to come about the
microscopic constituents of the universe would not be started
off moving randomly after all, but each little particle, each
electromagnetic wave, set off along a carefully chosen path to
lead to this very special future evolution. ... Such a changeover
requires ... an extraordinary degree of cooperation between countless
numbers of atoms.%
{\normalsize %
\footnote{Davies (1977), 195-6.}%
}

\end{list}%
\setlength{\parindent}{\defaultparindent}%
\setlength{\parskip}{0pt}%
Davies here alludes to his earlier conclusion that
`a time-asymmetric universe does not demand any very special
initial conditions. It seems to imply a creation which is of
a very general and random character at the microscopic level.'%
{\normalsize %
\footnote{Davies (1977), 193.}%
}%
{\normalsize  }However, we have seen that to maintain this view
of the early universe while invoking the usual statistical arguments
with respect to the late universe is to operate with a double
standard: double standards aside, GAS shows that if a late universe
is naturally clumpy, so too is an early universe. In the present
context the relevant point is that (as Davies himself notes,
in effect%
{\normalsize %
\footnote{Davies (1974), 199. }%
}) the conventional time-asymmetric view itself requires that
the final conditions of the universe be microscopically arranged
so that when viewed in the reverse of the ordinary sense, the
countless atoms cooperate over billions of years to achieve the
remarkable low entropy state of the Big Bang. Again, therefore,
a double standard is involved in taking it to be an argument
against Gold's view that it requires this cooperation in the
initial conditions. As before, the relevant statistical argument
is an instrument with two possible uses. We know that it yields
the wrong answer in one of these uses, in that it would exclude
an early universe of the kind we actually observe. Should we
take it to be reliable in its other use, which differs only in
temporal orientation from the case in which the argument so glaringly
fails? Symmetry and simple caution both suggest that we should
not!

\setlength{\parindent}{\defaultparindent}%
\setlength{\parskip}{0pt} A very different sort of objection
to the time-reversing model is raised by Penrose in the following
passage:

\begin{list}{ }{%
\setlength{\leftmargin}{35pt}\setlength{\rightmargin}{17.05pt}%
\setlength{\topsep}{0pt}\setlength{\partopsep}{0pt}}
\item

Let us envisage an astronaut in such a universe who falls into
a black hole. For definiteness, suppose that it is a hole of
10%
{\normalsize $^{10}$} [solar masses] so that our astronaut will
have something like a day inside [the event horizon], for most
of which time he will encounter no appreciable tidal forces and
during which he could conduct experiments in a leisurely way.
... Suppose that experiments are performed by the astronaut for
a period while he is inside the hole. The behaviour of his apparatus
(indeed, of the metabolic processes within his own body) is entirely
determined by conditions at the black hole's singularity ...---as,
equally, it is entirely determined by the conditions at the big
bang. The situation inside the black hole differs in no essential
respect from that at the late stages of a recollapsing universe.
If one's viewpoint is to link the local direction of time's arrow
directly to the expansion of the universe, then one must surely
be driven to expect that our astronaut's experiments will behave
in an entropy-%
{\it decreasing} way (with respect to `normal' time). Indeed,
one should presumably be driven to expect that the astronaut
would believe himself to be coming out of the hole rather than
falling in (assuming his metabolic processes could operate consistently
through such a drastic reversal of the normal progression of
entropy).%
{\normalsize %
\footnote{Penrose (1979), 598-9.}%
}

\end{list}%
\setlength{\parindent}{\defaultparindent}%
\setlength{\parskip}{0pt} This objection seems to me to put
unreasonable demands on the nature of the temporal reversal in
these time-symmetric models. Consider Penrose's astronaut. He
is presumably a product of 10%
{\normalsize $^{9}$} years of biological evolution, to say nothing
of the 10%
{\normalsize $^{10}$} years of cosmological evolution which created
the conditions for biology to begin on our planet. So he is the
sort of physical structure that could only exist at this kind
of distance from a suitable big bang. (What counts as %
{\it suitable}? The relevant point is that low entropy doesn't
seem to be enough; for a start, the bang will have to be massive
enough to produce the cosmological structure on which life depends.)

 All this means that our astronaut isn't going to encounter any
time-reversed humanoids inside the black hole, any unevolving
life, or even any unforming stars and galaxies. More importantly,
it means that he himself doesn't need to have an inverse evolutionary
history inside the hole, in addition to the history he already
has outside. He doesn't need to be a `natural' product of the
hole's singularity. Relative to its reversed time sense, he's
simply a miracle. The same goes for his apparatus---in general,
for all the `foreign' structure he imports into the hole.

 Notice here that there are two possible models of the connections
that might obtain between the products of two low entropy boundary
conditions: a `meeting' model, in which any piece of structure
or order is a `natural' product of both a past singularity and
a future singularity; and a `mixing' model, in which it is normally
a product of one or the other but not necessarily both. (See
Figure 1.) Penrose's argument appears to assume the meeting model.
Hawking also seems to assuming this model when he suggests%
{\normalsize %
\footnote{Hawking (1985), 2490.}%
} that the astronaut entering the event horizon of a black hole
wouldn't notice the time reversal because his psychological time
sense would reverse. As I say, however, this seems to me to place
a quite unnecessary constraint on the time-reversal view. The
appropriate guiding principle seems to be that any piece of structure
needs to explained %
{\it either} as a product of a past singularity %
{\it or} as a product of the future singularity; but that no
piece needs both sorts of explanation. The proportions of each
kind can be expected to vary from case to case. In may well be
that in our region of the universe, virtually all the structure
results from the Big Bang. This might continue to be the case
if in future we fall into the sort of black hole which doesn't
have the time or mass to produce much structure of its own. In
this case the experience might be very much less odd than Penrose's
thought experiment would have us believe. The reverse structure
produced by the black hole might be insignificant for most of
the time we survived within its event horizon.

\setlength{\parindent}{\defaultparindent}%
\setlength{\parskip}{0pt} What if we approach a black hole which
is big enough to produce interesting structure---the Big Crunch
itself, for example? Doesn't Penrose's argument still apply in
this case? It seems to me that the case is still far from conclusive,
so long as we bear in mind that %
{\it our} structure doesn't need a duplicate explanation from
the opposite point of view. It is true that in this case we will
expect eventually to be affected by the reverse structure we
encounter. For example, suppose that our spacecraft approaches
what from the reverse point of view is a normal star. From the
reverse point of view we are an object leaving the vicinity of
the star. We appear to be heated by radiation from the star,
but to be gradually cooling as we move further away from the
star, thus receiving less energy from it, and radiating energy
into empty space.

 What would this course of events look like from our own point
of view? Apparently we would begin to heat up as photons `inexplicably'
converged on us from empty space. This inflow of radiation would
increase with time. Perhaps even more puzzlingly, however, we
would notice that our craft was re-radiating towards one particular
direction in space---towards what from our point of view is a
giant radiation sink. Whether we could detect this radiation
directly is a nice question---more on this below---but we might
expect it to be detectable indirectly. For example we might expect
that the inside of the wall of the spaceship facing the reverse
star would feel cold to the touch, reflecting what in our time
sense would be a flow of heat energy towards the star.  These
phenomena would certainly be bizarre by our ordinary standards,
but it is not clear that their possibility constitutes an objection
to the possibility of entropy reversal. After all, within the
framework of the Gold entropy-reversing model itself they are
not in the least unexpected or inexplicable. To generate a substantial
objection to the model, it needs to be shown that it leads to
incoherencies of some kind, and not merely to the unexpected.
Whether this can be shown seems to be an open question. More
on this in the next section.

 Penrose himself no longer puts much weight on the astronaut
argument.%
{\normalsize %
\footnote{Though it appears as recently as Penrose (1989), 334-5.}%
} In recent correspondence he says that he now thinks that a
much stronger case can be made against the suggestion that entropy
decreases towards singularities. He argues that in virtue of
its commitment to temporal symmetry this view must either disallow
black holes in the future, or allow for a proliferation of white
holes in the past. He says that the first of these options `requires
physically unacceptable teleology', while the second would conflict
with the observed smoothness of the early universe.%
{\normalsize %
\footnote{Penrose (1991).}%
} However, the objection to the first option is primarily statistical:
`it would have to be a seemingly remarkably improbable set of
coincidences that would forbid black holes forming. The hypothesis
of black holes being not allowed in the future provides ``unreasonable''
constraints on what matter is allowed to do in the past.' %
{\normalsize %
\footnote{Penrose (1991).}%
} And this means that Penrose is again invoking the old double
standard, in accepting the `naturalness' argument with respect
to the future but not the past. Once again: the lesson of the
smooth past seems to be that in that case something overrides
the natural behaviour of a gravitational collapse; once this
possibility is admitted, however, we have no non-question-begging
grounds to exclude (or even to %
{\it doubt}!) the hypothesis that the same overriding factor
might operate in the future.%
{\normalsize %
\footnote{Cf. the argument in Davies (1974), 96, based on the
work of Rees. This too is effectively a statistical argument,
for the point is that `in the normal course of events' radiation
will not reconverge on stars. In the normal course of events
it wouldn't do so in the reverse direction either, but something
seems to override the statistical constraint.

Davies has another radiation-based argument against the Gold
universe: `Any photons that get across the switch-over unabsorbed
will find when they encounter matter that the prevailing thermodynamic
processes are such as to produce anti-damping .... If a light
wave were to encounter the surface of a metallic object, it would
multiply in energy exponentially instead of diminish. ... Consistency
problems of this sort are bound to arise when oppositely directed
regions of the universe are causally coupled together.' (Davies
(1974), 194) However, Davies apparently no longer regards this
as a powerful argument against the time-reversing view, for in
Davies and Twamley (1993) he and Jason Twamley canvas other ways
in which radiation might behave in a time-reversing cosmos. I
turn to the most significant of these arguments in the next section.}%
}

\begin{center}%
\setlength{\parindent}{\defaultparindent}Is a reversing future
observable? Is it consistent?

\end{center}Now to return to the issues we deferred above, concerning
the observability and hence the coherency of the Gold model.
To give the enquiry a slightly less speculative character, let
us locate our observers on Earth. Suppose that the actual universe
were a Gold universe, and a terrestrial telescope was pointed
in the direction of a reverse-galaxy---i.e., a galaxy in the
what in our time sense is the contracting half of the universe.
What effect, if any, would this have on the telescope and its
environs?

 As in the earlier case, a first step is to consider matters
from the reverse point of view. To an astronomer resident in
the reverse-galaxy, light emitted from that galaxy appears to
be being collected and eventually absorbed by the distant telescope
on Earth. This astronomer will therefore expect appropriate effects
to take place at the back of our telescope: a black plate there
will be heated slightly by the incoming radiation from the astronomer's
galaxy, for example. But what does this look like from our point
of view? Our temporal sense is the reverse of that of the distant
astronomer, so that what she regards as absorption of radiation
seems to us to be emission, and vice versa. Apparent directions
of heat flow are similarly reversed. Thus as we point our telescope
in the distant galaxy's direction, its influence should show
up directly in a suddenly increased flow of radiation from the
telescope into space, and indirectly as an apparent cooling in
the temperature of the black plate at the rear of our telescope.
The plate will apparently seem cooler than its surroundings,
in virtue of the inflow of heat energy which is to be re-radiated
towards the distant galaxy.

 As in normal astronomy, the size of these effects will naturally
depend on the distance and intensity of the reverse-source. Size
aside, however, there seem to be theoretical difficulties in
detecting the effects by what might seem the obvious methods.
For example, it will be no use placing a photographic plate over
the aperture of the telescope, hoping to record the emission
of the radiation concerned. If we consider things from the point
of view of the distant reverse-observer it is clear that the
plate will act as a shield, obscuring the telescope from the
light from this observer's galaxy. Thus from our point of view
the light will be emitted from the back of the plate: the side
facing away from the telescope, toward the reverse-galaxy.

 The important thing seems to be that the detector effect be
one that from the reverse point of view does not depend on any
very special initial state---since that would be a special final
state from our point of view, which accordingly we couldn't `prepare'.
Unfogging photo plates perhaps won't work for this reason, for
example. We can't guarantee that the plate finishes in an unfogged
condition (in the way that we can normally guarantee that it
starts that way). Provided this requirement for a `non-special'
boundary state is satisfied, we appear to be entitled to assume
that the detector will behave as we would expect it to if the
boot were on the other foot, so to speak: in other words, if
we were transmitting towards a detector belonging to the inhabitants
of the reversing-future.

 Let us now look at this behaviour in a little more detail. If
we shine a light at an absorbing surface we expect its temperature
to increase. If the incoming light intensity is constant the
temperature will soon stabilise at a higher level, as the system
reaches a new equilibrium. If the light is then turned off the
temperature then drops exponentially to its previous value. Hence
if future reverse-astronomers shine a laser beam in the direction
of one of our telescopes, at the back of which is an absorbing
plate, the temperature change they would expect to take place
in the plate is as shown in Figure~2a. When the telescope is
opened the temperature of the plate rises due to the effect of
the incoming radiation, stabilising at a new higher value. If
the telescope is then shut, so that the plate no longer absorbs
radiation, its temperature drops again to the initial value.

 Figure 2b shows what this behaviour looks like from our point
of view. As explained above, both the temporal ordering of events
and the direction of change of the apparent temperature of the
plate relative to its surroundings has to be reversed. One of
the striking things about this behaviour is that it appears to
involve advanced effects. The temperature falls before we open
the telescope, and rises before we close it. This suggests that
we might be able to argue that the whole possibility is incoherent,
using a version of the bilking argument. Couldn't we adopt the
following policy, for example: %
{\it Open the telescope only if the temperature of the black
plate has not just fallen significantly below that of its surroundings}.
It might seem that this entirely plausible policy forces the
hypothesis to yield contradictory predictions, thus providing
a %
{\it reductio ad absurdum} of the time-reversing view. However,
it is not clear the results are contradictory. Grant for the
moment that while this policy is in force it will not happen
that the temperature of the plate falls on an occasion on which
we might have opened the telescope, but didn't actually do so.
This leaves the possibility that on all relevant trials the temperature
does not fall, and the telescope is opened. Is this inconsistent
with the presence of radiation from the future reverse-source?

 I don't think it is. Bear in mind that the temperature profile
depicted in these diagrams relies on statistical reasoning: it
is inferred from the measured direction of heat flow, and simply
represents the most likely way for the temperature of the absorbing
plate to behave. But one of the lessons of our discussion has
been that statistics may be overridden by boundary conditions.
Here, the temperature is constant before the telescope is opened
because our policy has imposed this as a boundary condition.
A second boundary condition is provided by the presence of the
future reverse radiation source. Hence the system is statistically
constrained in both temporal directions. We should not be surprised
that it does not exhibit behaviour predicted under the supposition
that in one direction or other, it has its normal degrees of
freedom. It is not clear whether this loophole will always be
available, though my suspicion is that it will be. If nothing
else, quantum indeterminism is likely to imply that it is impossible
to sufficiently constrain the two boundary conditions to yield
an outright contradiction.

 A consistency objection of a different kind to the Gold universe
has been raised by Hartle and Gell-Mann.%
{\normalsize %
\footnote{Reported in Davies and Twamley (1993), p. 4 in preprint.}%
} Hartle and Gell-Mann point out that assuming the present epoch
is relatively `young' compared to the epoch of maximum expansion,
it is to be expected that the conversion of matter to radiation
in stars will raise the ratio of the energy density of radiation
to that of non-relativistic matter considerably by the time of
the corresponding epoch in the recollapsing phase. Hartle and
Gell-Mann estimate an increase by a factor of the order of 10%
{\normalsize $^{3}$}. By symmetry the same should apply in reverse,
so that at our epoch there should be much more radiation about
than is actually observed. (Indeed, the effect should be accentuated
by blue-shift due to the universe's contraction.) As Davies and
Twamley put it, `By symmetry this intense starlight background
should also be present at our epoch ..., a difficulty reminiscent
of Olbers' paradox.'%
{\normalsize %
\footnote{Davies and Twamley (1993), p. 4 in preprint.}%
}

\setlength{\parindent}{\defaultparindent}%
\setlength{\parskip}{0pt} One thing puzzles me about this argument:
if there were such additional radiation in our region, could
we detect it? After all, it is neatly arranged to converge on
its future sources, not on our eyes or instruments. Thus imagine
a reverse-source in direction %
{\it +x} is emitting (in its time-sense) towards a distant point
in direction %
{\it -x}. We stand at the origin, and look towards %
{\it -x}. (See Figure 3.) Do we see the light which in our time
sense is travelling from %
{\it -x} towards %
{\it +x}? No, because we are standing in the way! If we are
standing at the origin (at the relevant time) then the light
emitted from the reverse-galaxy falls on us, and never reaches
what we think of as the past sky. When we look towards %
{\it -x}, looking for the radiation converging on the reverse-galaxy
at %
{\it +x}, then the relevant part of the radiation doesn't come
from the sky in the direction %
{\it -x} at all; it comes from the surface at the origin which
faces %
{\it +x}---i.e., from the back of our head.

 The issue as to whether and to what extent the influence of
the low entropy condition of one end of a Gold universe might
be expected to be apparent at the other is clearly an important
one, if the Gold view is to be taken at all seriously. The main
lesson of these brief comments is that the issue is a great deal
more complicated than it may seem at first sight. The more general
lesson is that because our ordinary (asymmetric) habits of causal
and counterfactual reasoning are intimately tied up with the
thermodynamic asymmetry, we cannot assume that they will be
dependable
in contexts in which this asymmetry is not universal. To give
a crude example, suppose that an event B follows deterministically
from an event A. In a Gold universe we may not be able to say
that if A had not happened B would not have happened; not because
there is some alternative earlier cause waiting in the wings
should A fail to materialise (as happens in cases of pre-emption,
for example), but simply because B is guaranteed by %
{\it later} events.

 Indeed, Figure 3 illustrates a consequence of this kind. We
had a choice as to whether to interpose our head and hence our
eye at the point %
{\it O}. Had we not done so, the light emitted (in the reverse
time sense) by the reverse-galaxy at %
{\it +x} would have reached %
{\it -x}, %
{\it in our past}. Our action thus influences the past. Because
we interpose ourselves at %
{\it O}, some photons are not emitted from some surface at %
{\it -x}, whereas otherwise they would have been. Normally claims
to affect the past give rise to causal loops, and hence inconsistencies.
But again it doesn't seem to me to be obvious that this will
happen in this case, for reasons similar to those in the telescope
case.

 These issues clearly require a great deal more thought. Let
me therefore conclude this section with two rather tentative
claims. First, it has not been shown that the reversing-universe
view leads to incoherencies. And second, there seems to be some
prospect that the contents of a time-reversing future universe
might be presently observable, at least in principle.%
{\normalsize %
\footnote{The required experiment should not be confused with
a well-known test of some cosmological implications of the Wheeler-
Feynman
absorber theory of radiation. A prediction of that theory was
that a transmitter should not radiate at full strength in directions in which
the future universe is transparent to radiation. Partridge (1973)
performed a version of this experiment with negative result.
The present discussion does not depend on the absorber theory
(which seems to me misconceived; see Price (1991)),
and predicts a different result, namely increased radiation in
the direction of future reverse-sources.}%
} The methods involved certainly look bizarre by ordinary standards;
but in the end this is nothing more than the apparent oddity
of perfectly ordinary asymmetries having the reverse of their
`usual' orientation. And the main lesson of this paper is that
unless we have learnt to disregard that sort of oddity, we won't
get anywhere with the problem of explaining temporal asymmetry.

\begin{center}%
\setlength{\parindent}{\defaultparindent}%
\setlength{\parskip}{0pt}Conclusions

\end{center}What then are the options and prospects for an explanation
of the observed cosmological time-asymmetry? Let us try to summarise.

 It might seem that the most attractive solution would be the
possibility mentioned in our discussion of Hawking's NBC, namely
a demonstration that although the laws that govern the universe
are temporally symmetric, the universes that they permit are
mostly asymmetric---mostly such that they possess a single temporal
extremity with the ordered characteristics of what we call the
Big Bang. But it cannot be over-emphasised that the usual statistical
considerations %
{\it do not} make this solution intrinsically more likely or
more plausible than the Gold time-symmetric cosmology. With double
standards disallowed, the statistical arguments concerned are
simply incompatible with the hypothesis that the Big Bang itself
is explicable as anything more than a statistical fluke. So if
anything it is the Gold view which should be regarded as the
more plausible option, simply on symmetry grounds---at least
in the absence of properly motivated consistency objections to
time-reversing cosmologies. And failing either of these approaches,
the main option seems to be an anthropic account. True, there
is also Penrose's view, but here the asymmetry is rather %
{\it ad hoc}. If we are going to invoke an %
{\it ad hoc} principle, we might as well have a symmetric one,
at least in the absence of decisive objections to the Gold cosmology.

 However, it should be emphasised that none of these alternatives
is immediately attractive. As we saw, the anthropic approach
involves an enormous ontological cost---it requires that the
universe be vastly larger than what we know as the observable
universe. (If we take Penrose's calculation as an indication
of how unlikely the observed universe actually is, and assume
that it is more or less as likely as it can be, consistent with
the presence of observers, then we have an estimate of the size
of this ontological cost: our universe represents at best 1 part
in 10$^{10^{30}}$ of the whole thing.) And as for the Gold
universe, it is no magic solution to the original problem, even
if consistent. We still need to explain why singularities are
of low entropy. Unless it can actually be shown that a generic
gravitational singularity is of this kind, some additional boundary
constraint will again be required. Although in the Gold case
this additional constraint need not be time-asymmetric, it may
still appear somewhat %
{\it ad hoc}. As noted earlier, however, this might seem preferable
to having no explanation of what would otherwise seem such a
striking physical anomaly. Explanatory power and theoretical
elegance are the traditional antidotes to apparent %
{\it ad hoccery}. A proposal of sufficient formal merit---perhaps
the original symmetric version of Hawking's NBC, for example---might
seem a very satisfactory theoretical solution.

 What would have to be established to avoid the need for such
an additional boundary constraint altogether? To answer this,
recall that the heart of GAS was the observation that any gravitating
universe still looks like a gravitating universe when its temporal
orientation is reversed. It follows that if increasing entropy
were the inevitable result of gravitational relaxation, any gravitating
universe would look like an entropy-maximising universe %
{\it from both temporal viewpoints}. It was the inconsistency
between this conclusion and the observed the low entropy Big
Bang that showed us that statistical reasoning is untrustworthy
in this cosmological context---that undercut our grounds for
thinking that gravitational relaxation inevitably or `normally'
produces an increase in entropy.

 I want to close by noting that in principle there seem to be
two ways to reconcile statistics and low entropy boundary conditions.
(I don't mean to suggest that either of these alternatives should
necessarily be taken seriously; I am simply interested in sketching
the logical structure of the problem.) One way would be to make
cosmological entropy, like gravitation, a viewpoint-dependent
matter. We have already noted that in whichever temporal orientation
one regards the universe, it appears to be gravitating---i.e.
subject to the influence of an attractive gravitational force.
It is therefore an orientation-dependent matter as to which temporal
extremity we take to involve the state of greatest gravitational
relaxation (or lowest gravitational potential energy). One way
to preserve a strong link between entropy and gravitation would
be to suggest that entropy is similarly frame-dependent, and
hence that entropy appears to increase monotonically from both
temporal viewpoints.

 The second way would be show that gravitational collapse does
not naturally lead to a high entropy singularity at all---in
other words, to find within one's theory of gravity an argument
to the effect that entropy naturally %
{\it decreases} in gravitational collapse. In this connection
it is interesting to note a recent paper by Sikkima and Israel%
{\normalsize %
\footnote{Sikkima and Israel (1991). }%
} claiming to show that a low entropy state may indeed be the
`natural' result of gravitational collapse. In one sense this
seems to be just the sort of thing that would have to be established,
if the boundary conditions are not to require an independent
law. However, Sikkima and Israel do not see the argument as supporting
the time-symmetric view. For one thing, they say that in a cyclical
universe entropy will increase from cycle to cycle. So the old
puzzle would re-emerge at this level: How can such an overall
asymmetry be derived from symmetric assumptions?

\setlength{\parindent}{\defaultparindent}%
\setlength{\parskip}{0pt} The two approaches just described
sought to reconcile statistical arguments in cosmology with the
low entropy Big Bang. An alternative would be to dispense with
the statistical considerations altogether. It might be argued
that any such appeal to statistics relies on the assumption that
the system in question has the freedom to choose its path from
among a range of equally likely futures. Perhaps the root of
our dilemma simply lies in this assumption---in the fact that
the assumption is itself incompatible with the neutral atemporal
perspective we must adopt if we are to explain temporal asymmetry.
That is, perhaps we should restore the balance not by seeking
a `natural' endpoint in both directions, but by curing ourselves
of our lingering attachment to the idea of natural evolution
itself. Perhaps in thinking that low entropy endpoints are anomalous
we have been misled by a form of reasoning which is itself grounded
in the temporal asymmetry; hence a form of reasoning we should
be prepared to discard, in moving to an atemporal viewpoint.
This would be a more radical departure from our ordinary ways
of thinking than anything contemplated in this paper. All the
same, a possible conclusion seems to be that the project of explaining
temporal asymmetry is a hopeless one, unless we are prepared
to contemplate a radical departure of some such kind.

\vskip 1cm

\begin{center}%
Acknowledgements

\end{center}This paper was originally written for the {\it Time's Arrows
Today} conference at UBC, Vancouver, in June 1992. I would
like to thank Steve Savitt for his comments and encouragement. I am
grateful to participants in the conference and to two later audiences in
Sydney for their help
in clarifying these ideas. I am also indebted to Dieter Zeh for patient
correspondence
on these issues over a number of years, and to Professor Roger
Penrose for the correspondence mentioned in footnote 41.

\newpage
\begin{center}
References
\end{center}

\setlength{\parindent}{\defaultparindent}
Davies, P. C. W. (1974) %
{\it The Physics of Time Asymmetry}. London: Surrey University
Press.

Davies, P. C. W. (1977) %
{\it Space and Time in the Modern Universe}. Cambridge: Cambridge
University Press.

Davies, P. C. W. (1983). `Inflation and Time Asymmetry in the
Universe', %
{\it Nature}, %
{\bf 301}, 398-400.

Davies, P. C. W. and Twamley, J. (1993). `Time-Symmetric Cosmology
and the Opacity of the Future Light Cone', %
{\it Classical and Quantum Gravity}, %
{\bf 10}, 931-45.

Gold, T. (1962). `The Arrow of Time', %
{\it American Journal of Physics}, %
{\bf 30}, 403-410.

Halliwell, J., Perez-Mercader, J. and Zurek, W. (eds.) (1993).
{\it Physical Origins of Time Asymmetry: Proceedings of the
NATO Workshop in Mazagon, Spain, October, 1991}. Cambridge:
Cambridge
University Press.

Hawking, S. W. (1985). `Arrow of Time in Cosmology', %
{\it Physical Review D}, %
{\bf 33}, 2489-95.

Hawking, S. W. (1988) %
{\it A Brief History of Time}. London: Bantam.

Hawking, S. W. (1993) `The No Boundary Condition and the Arrow
of Time'. In Halliwell, J., Perez-Mercader, J. and Zurek, W.
(1993).

Hawking, S. W. and Halliwell, J. (1985). `Origins of Structure
in the Universe', %
{\it Physical Review D}, %
{\bf 31}, 1777-91.

Hawking, S. W. and Israel, W. (eds.) (1979). %
{\it General Relativity: an Einstein Centenary}. Cambridge:
Cambridge University Press.

Hawking, S. W. and Israel, W. (eds.) (1987). %
{\it Three Hundred Years of Gravitation}. Cambridge: Cambridge
University Press.

Lewis, D. (1986) %
{\it The Plurality of Worlds}. Oxford: Blackwell.

Lind\'{e}, A. (1987) `Inflation and Quantum Cosmology'. In Hawking,
S. W. and Israel, W. (1987): 604-630.

Page, D. N. (1983). `Inflation Does Not Explain Time Asymmetry',
{\it Nature}, %
{\bf 304}, 39-41.

Partridge, R. B. (1973). `Absorber Theory of Radiation and the
Future of the Universe', %
{\it Nature}, %
{\bf 244}, 263-265.

Penrose, R. (1979) `Singularities and Time-Asymmetry'. In Hawking,
S. W. and Israel, W. (1979): 581-638.

Penrose, R. (1989) %
{\it The Emperor's New Mind}. Oxford: Oxford University Press.

Penrose, R. (1991). Personal Correspondence, 28.21.91 \& 21.2.91.

Price, H. (1989). `A Point on the Arrow of Time', %
{\it Nature}, %
{\bf 340}, 181-2.

Price, H. (1991). `The Asymmetry of Radiation: Reinterpreting
the Wheeler-Feynman Argument', %
{\it Foundations of Physics}, %
{\bf 21}, 959-975.

Sikkima, A. E. and Israel, W. (1991). `Black-hole Mergers and
Mass Inflation in a Bouncing Universe', %
{\it Nature}, %
{\bf 349}, 45-47.

\end{document}